\documentstyle[12pt,eqsecnum,preprint,prd,aps,epsf]{revtex}
\begin{document}
\draft
\preprint{ \hbox{FTUAM 95--32} }
\date{September, 1995}
\title{
Improved Evaluation of the Hadronic Vacuum Polarization
Contributions to Muon $g-2$ and $\bar\alpha_{\rm QED}(M_Z)$
Using High Order QCD Calculations\footnote{Research partially
supported by CICYT,Spain.}}
% \footnote{Electronic address: {\tt
% kassa@nantes.ft.uam.es.}}
\author{K.~Adel and F.~J.~Yndur\'ain\\
Departamento de F\'{\i}sica Te\'orica, C-XI,\\ Universidad
Aut\'onoma de Madrid, Canto Blanco,\\ E--28049 Madrid,
SPAIN}
\maketitle
\begin{abstract}
We use recently evaluated radiative and nonperturbative corrections to
production of heavy quarks by a vector current to give very precise
theoretical calculations of the high energy ($t^{1/2}\geq \sqrt{2}$
GeV) imaginary part of the photon vacuum polarization function, ${\rm
Im}\Pi(t)$.
This allows us to improve the corresponding contributions to the muon
(or any other lepton) $g-2$ anomaly and to the running QED constant on
the $Z$, $\bar\alpha_{\rm QED}(M_Z)$. This decreases the error in the
evaluations by a factor between two and six for the high energy
contribution, and by some 50\% for the overall result.
We find for the hadronic contributions
$a_h=6993.4\pm110.0\times10^{-11}$ and
$\Delta\alpha_h=272.59\pm4.09\times10^{-4}\,.$
\end{abstract}

% %%%%%%%%%%%%%%%%%%%%%%%%%%%%%%%%%%%%%%%%%%%%%%%%%%%%%%%%%%%%%%%%%%%%%%%%
\section{Introduction}
\label{sec:1}
% %%%%%%%%%%%%%%%%%%%%%%%%%%%%%%%%%%%%%%%%%%%%%%%%%%%%%%%%%%%%%%%%%%%%%%%%

There has been recently a renewed interest$^{\cite{bb:eidelman}}$
on two basic quantities,
both related to the photon vacuum polarization: the hadronic
contribution to the anomalous moment of the muon, $a_{\mu}$, and the
running QED charge $\overline{\alpha}$ defined at the $Z$ particle
pole. In what respects the first, the reason is the advanced status
of the experiment planned at Brookhaven which should improve the
present accuracy by more than one order of magnitude; in what respects
the second, because the quantity $\overline{\alpha}_{QED}(M_Z^2) $
plays a leading role in precision determination of electroweak
parameters.

In this paper we consider the contributions of the hadronic part of the
photon vacuum polarization tensor to $a_{\mu}$ and
$\overline{\alpha}_{QED}(M_Z^2)$, that we write respectively as
\begin{equation}
a_{\mu}^{(2)}({\rm had}) \ \ \ \ \ ,\ \ \ \ \
\Delta \overline{\alpha}_{QED}^{(2)}({\rm had} \;,\;M_Z^2)\ .
\eqnum{1.1}
\end{equation}
These quantities may be expressed in terms of the photon hadron vacuum
polarization function $\Pi_{h}$.  One can then write a dispersion
relation for this function in such a way that the experimentally
accessible quantity
\begin{equation}
R(t) = { \sigma (e^+ e^- \rightarrow \gamma^* \rightarrow hadrons)
\over \sigma (e^+ e^- \rightarrow \mu^+ \mu^-) }
\eqnum {1.2}
\end{equation}
appears instead of $\Pi_h$:
\begin{eqnarray}
\eqnum{1.3}
&& a_h \equiv
 a_{\mu}^{(2)}({\rm had}) =
 \int_{4\,m_\pi^2}^{\infty} dt\; R(t) \; K(t) \ \ ,\\
\eqnum{1.4}
&& \Delta \alpha_h \equiv
\Delta \overline{\alpha}_{QED}^{(2)}({\rm had} \;,\;M_Z^2)
	= - {\alpha_{QED}\,M_Z^2 \over 3\,\pi }\;
   \int_{4\,m_\pi^2}^{\infty} dt\; {R(t) \over t(t-M_Z^2)} \ \ .
\end{eqnarray}
Here,
\begin{equation}
\eqnum{1.5}
K(t) = {\alpha_{QED} \over 3\,\pi^2 \,t}\;
	\int_{0}^{1} dz \;{ z^2(1-z) \over z^2 + (1-z)\,t/m_{\mu}^2 }\ .
\end{equation}

The reason why we think we can improve on existing estimates of
$a_h,\ \Delta \alpha_h$ is that we may use the very reliable {\em
theoretical} QCD calculations\footnote{The fact that QCD calculations
are more precise than the use of experimental data was already noted
in e.g. Ref. \cite{bb:casas}.}
that have been extended with great precision to
regions previously unaccessible; in particular$^{\cite{bb:kassa}}$
to the regions
right above threshold for heavy quark production ($c\bar{c},\
b\bar{b}$) as well as the very low ($t^{1/2} \sim 1.2 \;GeV$)
energies.
The extensions have been possible because of
the high orders attained by the QCD
calculations$^{\cite{bb:chetyrkin}}$
and the use of the precisely known value of
$\alpha_s$ on the $\tau$ mass$^{\cite{bb:pich}}$,
\begin{equation}
\eqnum{1.6}
\alpha_s( m_{\tau}^2 ) = 0.33 \pm 0.05 \ \ .
\end{equation}
Likewise, the use of the value$^{\cite{bb:pdb}}$ of $\alpha_s$ at the $Z$,
\begin{equation}
\eqnum{1.7}
\alpha_s( M_Z^2 ) = 0.119 \pm 0.003 \ \ ,
\end{equation}
allows us to reduce still further the uncertainties in the high energy
regions.

% %%%%%%%%%%%%%%%%%%%%%%%%%%%%%%%%%%%%%%%%%%%%%%%%%%%%%%%%%%%%%%%%%%%%%%%%
\section{Contribution from the region
$\ \lowercase{t}^{1/2}<1.1\;G\lowercase{e}V$ }
\label{sec:2}
% %%%%%%%%%%%%%%%%%%%%%%%%%%%%%%%%%%%%%%%%%%%%%%%%%%%%%%%%%%%%%%%%%%%%%%%%

In the region $t^{1/2} < 1.1\;GeV$, perturbative QCD is clearly invalid.
One thus has to rely on experiment, supplemented by old-fashioned
hadron theory. The region may be further split into the $\rho$
resonance region (say, $t^{1/2} \le 0.8\; GeV$) and the rest. For
$a_h$ we have,
\begin{equation}
a_h(t^{1/2} < 0.8 ) = ( 4821 \pm 24 \pm 27 )\times 10^{-11} \;
+ (25 \pm 3)\times 10^{-11} \eqnum{2.1}\ .
\end{equation}
Here the second term in the r.h.s. is the $\omega-\rho$ interference
contribution. The first error is statistical; the second (when given)
will be the systematic one. Eq. (2.1) presents the value reported in Ref.
\cite{bb:casas};
other authors give compatible estimates.
We will improve on $(2.1)$ slightly later on.

To this one has to add the contribution of the region
$0.8 < t^{1/2} \le 1.1 $ which gives, using experimental data only,
\begin{equation}
a_h (0.8 < t^{1/2} \le 1.1) = (1100 \pm 94) \times 10^{-11} \ .\eqnum{2.2}
\end{equation}
This region, about which little can be done at present, presents
the largest source of error. Combining
Eqs. (2.1) and (2.2) we have,
\begin{equation}
a_h ( t^{1/2} \le 1.1 \; GeV) = (5947 \pm 97 \pm 27) \times 10^{-11} \ .
\eqnum{2.3}
\end{equation}

For $\Delta\alpha_h$, we have, taking the analysis of the first
article in Ref. \cite{bb:eidelman},
\begin{equation}
 \Delta \alpha_h ( t^{1/2} < 1.1\; GeV) =
	(35 \pm 0.3 \pm 0.7) \times 10^{-4}\ .
\eqnum{2.4}
\end{equation}

The contribution of the region
$\ 1.1 < \lowercase{t}^{1/2}< 2 \;G\lowercase{e}V$
is the topic of next section.

% %%%%%%%%%%%%%%%%%%%%%%%%%%%%%%%%%%%%%%%%%%%%%%%%%%%%%%%%%%%%%%%%%%%%%%%%
\section{Contribution from the region
$\ 1.1 < \lowercase{t}^{1/2}< 2 \;G\lowercase{e}V$ }
\label{sec:3}
% %%%%%%%%%%%%%%%%%%%%%%%%%%%%%%%%%%%%%%%%%%%%%%%%%%%%%%%%%%%%%%%%%%%%%%%%

{}From a theoretical point of view, the contribution of this region
presents a challenge, because one cannot use perturbative QCD
reliably, and also because old-fashioned hadron theory does not
describe very well the average experimental data points. In this section,
we explain how one can combine these two theories and give
reliable {\em theoretical} estimates, which can then be compared with
previous estimates based on {\em experimental} data
only$^{\cite{bb:eidelman}}$:
\begin{eqnarray}
\eqnum{3.1}
&& a_h^{\rm {exp}} ( 1.1 \le t^{1/2} < \sqrt{2} \; GeV) =
(278 \pm 25) \times 10^{-11} \ \ ,\\
\eqnum{3.2}
&& \Delta \alpha_h^{\rm {exp}} ( 1.1 < t^{1/2} < \sqrt{2}\; GeV) =
	(13 \pm 0.15 \pm 0.8) \times 10^{-4}\ .
\end{eqnarray}

Let us first estimate the contribution of this region to $R(t)$ using
perturbative QCD only: $R(t)$ may be split as a sum over light quark
flavors:
\begin{equation}
\eqnum{3.3}
R(t) = \sum_{q=u,d,s} R_q(t)\ \ ,
\end{equation}
and one may use the very precise high energy result for $R_q(t)$:
\begin{eqnarray}
\nonumber
R_q^{\rm{h.e.}}(t) \ {\mathop =_{ \bar{v}\to 1}} \ N_c \, Q_q^2\,&&\left\{
 1-{3\over 2}(1-{{\bar{v}}})^2 + {1\over 2}(1-{{\bar{v}}})^3 + \left[
{3\over 4} + {9\over 2} (1-{{\bar{v}}}) \right]
\;{{C_{\!_F}} \alpha_{\hskip-1.pt s}\over \pi}
\right.\\
\nonumber
&& \left.
+ \left[ {9\over 2} \left( \ln{2\over
1-{{\bar{v}}}} -{3\over 8} \right)(1-{{\bar{v}}})^2 \right]
{{C_{\!_F}} \alpha_{\hskip-1.pt s}\over \pi}
+ r_2 \, \left( {\alpha_{\hskip-1.pt s} \over \pi }\right)^2
+ \widetilde{r}_3 \, \left( {\alpha_{\hskip-1.pt s} \over \pi }\right)^3
\right.\\
\eqnum {3.4}
&& \
\left.
+ {9\over 2}(1-\bar{v}) {{C_{\!_F}}}
	\left[
		8.7 \,\left({\alpha_{\hskip-1.pt s} \over \pi } \right)^2
		+ 45.3 \,\left({\alpha_{\hskip-1.pt s} \over \pi }\right)^3
	\right]\;\right\}.
\end{eqnarray}
Here,
\begin{eqnarray}
\nonumber
&& {{\overline{v}}} \equiv
 \sqrt{ 1 - 4\,{{\overline{m}}}^2(t)/t } \ \ \ \ , \ \ \ \ \
   r_2 = 1.986 -0.115\,n_f \ \ \ , \ \ C_{_F}=4/3 \ ,\\
\eqnum{3.5}
&& \widetilde{r}_3 = -6.637 -1.2\,n_f -0.005\,n_f^2
- 1.24 \left( \sum_f Q_f \right)^2 \ \ \ ,
\end{eqnarray}
and $\overline{m}(t)$ and $\alpha_s\equiv \alpha_s(t)$
are the running mass and coupling constant which we may take to two
loops$^{\cite{bb:yndurain}}$.
%
% \begin{eqnarray}
% \eqnum{3.6}
% &&
% \overline{m}(t) = \widehat{m}
% \left[ {{\beta_{_0}} \alpha_s(t) \over 2\,\pi } \right]^{-{\gamma_{_0}}
%%/{\beta_{_0}} }
% \left\{ 1 + \left( - {{\gamma_{_1}} \over {\beta_{_0}}}
% + {\beta_{_1} {\gamma_{_0}} \over {\beta_{_0}}^2} \right) \,
%%{\alpha_s(t)\over \pi}
%   \right\} \ ,\\
% &&
% \nonumber
% \gamma_0 = - 3\,{C_{\!_F}} \ \ \ ,
%  \ \ \gamma_1 = - {3\,{C_{\!_F}}^2 \over 2}
% 	- {97\,{C_{\!_F}}{C_{\!_A}} \over 6} + { 5\,{C_{\!_F}} n_f \over 3},\\
% %
% \nonumber
% && {C_{\!_A}}=3\ , \ {T_{\!_F}}={1 \over 2} \ , \ C_{\!_F} = {4\over 3} \ ,
% \ {\beta_{\!_0}} = {11 \,{C_{\!_A}} - 4 \,{T_{\!_F}} \,n_f \over 3 } \ \ ,
% \ \ {\beta_{\!_1}} = 102 - 38\,n_f/3\ .
% %
% \end{eqnarray}
% %
%
The {\em justification} of this approximation lies in the fact
that the light quarks are relativistic at these
energies. Numerically, we take $\overline{m}_u = \overline{m}_d = 0$,
and use$^{\cite{bb:yndurain}}$
\begin{equation}
\eqnum{3.6}
\overline{m}_s(1\;GeV^2) = 0.19 \ GeV \ .
\end{equation}
The result $R_{QCD}(t)$ obtained in this approximation is
plotted in Fig. 1.  together with some experimental points. Although
the experimental errors are large, it is clear that our curve
$R_{QCD}$ will not give a reliable estimate.  Before we explain how
this can be improved, let us first describe another approximation
for $R(t)$ based on old-fashioned hadron theory.

\vspace{5mm}

\begin{center}
    \leavevmode
    \epsfverbosetrue
    \epsfxsize=3.5truein
    \epsffile{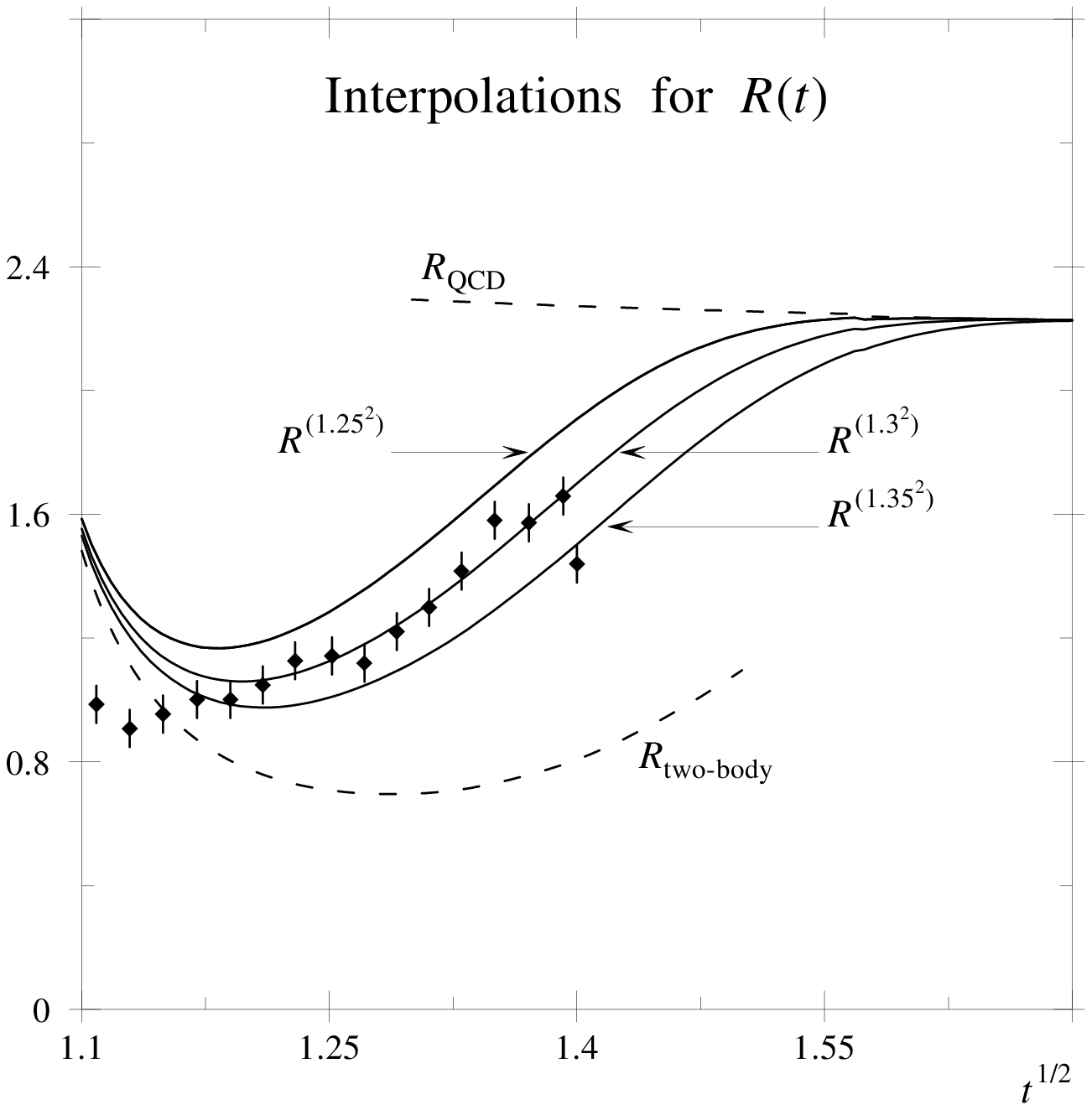}
\end{center}

\begin{center}
{\bf Fig. 1.}
\end{center}

To do this, we evaluate the contribution of the individual
channels, in the quasi-two-body approximation.
Thus, we have the two-body channels $\pi^+\pi^-$, $K^+K^-$,
$K^0\overline{K}^0$,
and the quasi-two-body channels
$\rho^{+} \pi^{-} $,
$\rho^{-} \pi^{+} $,
$\rho^{0} \pi^{0}$,
$\rho^{0} \eta$,
$\omega \pi^{0}$,
$\phi^{0} \pi^{0}$,
$\phi^{0} \eta$,
$K^{*\,0} K^{0}$,
$K^{*\,+} K^{-}$,
$K^{*\,-} K^{+} $.

The contribution of the first is estimated using the corresponding
form factors,
$\rho$ or $\phi$ dominated: we obtain
\begin{equation}
\Delta R(t)^{\rm{resonances}} =
\left| F_{\pi}(t) \right|^2 {1\over 4}\left( 1- {4\,m_{\pi}^2 \over t} \right)
+
\left| F_{K}(t) \right|^2 {1\over 4}\left( 1- {4\,m_{K}^2 \over t} \right)
\ \ , \eqnum{3.7}
\end{equation}
where
\begin{eqnarray}
\nonumber
F_{\pi}(t) &&= { m_{\rho}^2 + m_{\rho} m_{\pi} \xi_0 \over
m_{\rho}^2 - t - i\,m_{\rho} (t/4-m_{\pi}^2)^{3/2} \,\xi_0 }
\hskip1truecm , \hskip1truecm
\xi_0 \equiv {\Gamma_{\rho} \over (m_{\rho}^2/4 - m_{\pi}^2 )^{3/2} }\ ;\\
\nonumber
F_{K}(t) &&= { m_{\phi}^2 + m_{\phi} m_{K} \xi_1 \over
m_{\phi}^2 - t - i\,m_{\phi} (t/4-m_{K}^2)^{3/2} \,\xi_1 }
\hskip1truecm , \hskip1truecm
\xi_1 \equiv {\Gamma_{\phi} \over (m_{\phi}^2/4 - m_{K}^2 )^{3/2} }\ .
\end{eqnarray}
Note that we have combined the contribution coming
from both $K^+K^-$ and $K^0\overline{K}^0$ into the function
$F_K$ .

The quasi-two-body channels contribution to $R(t)$ is estimated in the
vector-meson dominance approximation:
\begin{equation}
\eqnum{3.8}
\Delta R(t)^{\rm{quasi-2-body}} =
 \sum_{{V,P}=\rho+\pi^-,\dots}{\Gamma(\gamma^* \rightarrow V P)
\over \Gamma(\gamma^* \rightarrow \mu^+\mu^-) }\ \ .
\end{equation}
Using the phenomenological interactions
\begin{equation}
{\cal{L}}_{VP} = g_{VP} \,\epsilon^{\mu\nu\alpha\beta} F_{\mu\nu}
V_{\alpha\beta}\phi_{P}\ \ ,
\eqnum{3.9}
\end{equation}
one can relate this to the radiative decay widths,
\begin{equation}
\Delta R(t)=  \sum_{{V,P}=\rho+\pi^-,\dots}
{3\, m_V^3 \, \Gamma(V \rightarrow \gamma \,P) \over
\alpha_{QED}\, (m_V^2-m_P^2)^3 }
{ \left[ t - (m_V+m_P)^2 \right]^{3/2} \left[ t - (m_V-m_P)^2 \right]^{3/2}
 \over t^{1/2} \, (t-4\,m_{\mu}^2)^{3/2} }\ .
\eqnum{3.10}
\end{equation}
The decay widths are taken from experiment.
The result $R(t)_{\rm{two-body}}$ we obtain in this approximation is
also plotted in Fig. 1. It unfortunately does
not represent an average of the experimental points.  To bring these
approximations closer to experiment, we construct the following
interpolation curve which reproduces
both evaluations in their ranges of validity: we note note that QCD
works at the higher energies, and the old-fashioned hadron evaluation
at the lower energies. Then,
\begin{eqnarray}
\nonumber
&& R^{(t_0)}(t) \equiv R(t)_{\rm{two-body}} \; \exp( - t^5/\sqrt{5}\,t_0^5 )
+ R_{QCD}(t) \;\left[ 1 - \exp( - t^5/\sqrt{5}\,t_0^5 ) \right] \ ,\\
\eqnum{3.11}
&& \ \ \ t_0^{1/2} = 1.25\ ,\ 1.3 \ ,\ 1.35 \ GeV \ \ .
\end{eqnarray}
The point $t_0$ is chosen to fit experiment.
These curves are drawn in Fig. 1 (solid curves) and represent
a significant improvement from what we had previously.
The numerical results obtained with these interpolation functions
are:
\begin{eqnarray}
\eqnum{3.12}
&& a_h ( 1.1 < t^{1/2} \le \sqrt{2} \; GeV) =
(265 \pm 22) \times 10^{-11} \ ,\\
\eqnum{3.13}
&& \Delta \alpha_h ( 1.1 < t^{1/2} \le \sqrt{2} \; GeV ) =
	(15.3 \pm 1.4) \times 10^{-4}\ .
\end{eqnarray}
The central values have been obtained using $R^{(t_0^{1/2}=1.3\;
GeV)}$, and the systematic errors have been estimated from the
difference $\ \left|R^{(t_0^{1/2}=1.3\;GeV)} -
R^{(t_0^{1/2}=1.35\;GeV)}\right|$. (These errors are in our opinion
over-estimated and should in principle be
reduced. We will however not do that here).
Comparing the numbers given in Eqs. (3.12, 3.13) with those given in
Eqs. (3.1, 3.2) we see they agree, within errors. We however believe
our evaluation to be more reliable than those estimates
which were obtained using experimental data only, because of the known
existence of large systematic errors of the last.

% %%%%%%%%%%%%%%%%%%%%%%%%%%%%%%%%%%%%%%%%%%%%%%%%%%%%%%%%%%%%%%%%%%%%%%%%
\section{Contribution from the region
$\ \lowercase{t} > 2\;G\lowercase{e}V^2$}
\label{sec:4}
% %%%%%%%%%%%%%%%%%%%%%%%%%%%%%%%%%%%%%%%%%%%%%%%%%%%%%%%%%%%%%%%%%%%%%%%%

For $t>2\;GeV^2$, we distinguish two different contributions. The first
is due to the $J/\psi$ and $\Upsilon$ bound states and the results can be
taken from the first article of Ref. \cite{bb:eidelman}:
\begin{eqnarray}
\eqnum{4.1}
&& a_h ( J/\psi)   = (86 \pm 4.1 \pm 4) \times 10^{-11} \ ,\\
\eqnum{4.2}
&& a_h ( \Upsilon) = (1 \pm 0 \pm 0.1) \times 10^{-11} \ ,\\[5pt]
\eqnum{4.3}
&& \Delta \alpha_h  ( J/\psi) =
	(11.34 \pm 0.55 \pm 0.61) \times 10^{-4}\ ,\\
\eqnum{4.4}
&& \Delta \alpha_h  ( \Upsilon) =
	(1.18 \pm 0.05 \pm 0.06) \times 10^{-4}\ .
\end{eqnarray}

The second contribution comes from the continuum regions, i.e. the
regions above $q\bar{q}$ thresholds. As before, we split $R(t)$ as
\begin{equation}
\eqnum{4.5}
R(t) = \sum_{q=u,d,s,c,b,t} R_q(t)\ \ .
\end{equation}
For the light quarks $u,\;d,\;s$, only the high energy region needs to
be considered; the corresponding expression for $R_q$ was given in Eq.
(3.4). For the {\em heavy} quarks, $c,\ b,\ t,$ we also need the value
of $R_q$ for small and intermediate velocities.  At low energy, the
value of $R_q$ is known with great precision$^{\cite{bb:kassa}}$:
\begin{eqnarray}
\nonumber
R_q^{\rm{l.e.}}(t) \ {\mathop =_{ v\to 0}} \ N_c \, Q_q^2\,&&\left\{
{v(3-v^2) \over 2} + \left( - {6\,v\over \pi} + {3\,\pi\,v^2 \over 4
}\right) \,{C_{\!_F}}\,\widetilde{\alpha}_{\hskip-1.3pt s} \right\}
\left( 1 - {2\,\pi \langle \alpha_{\hskip-1.pt s} \,G^2\rangle \over
192 \, m^4\,v^6} \right) \\
\eqnum {4.6}
&& \times \big[ 1+2\,c_{\,0}(t) \big] \, { \pi\,{C_{\!_F}}
\widetilde{\alpha}_{\hskip-1.3pt s} /v \over 1 -
e^{-\pi{C_{\!_F}}\widetilde{\alpha}_{\hskip-1.3pt s}/v } }\ \ ,\\
\nonumber
&& v \equiv \sqrt{1-4\,m^2/t} \ \ \ \ , \ \ \ \ \
\widetilde{\alpha}_{\hskip-1.3pt s} = \alpha_{\hskip-1.pt s}(t) \;
\left[ 1 + { a_1 + {\gamma_{\!_E}} {\beta_{\!_0}}/2 \over \pi } \;
	\alpha_{\hskip-1.pt s}(t) \right] \ ,\\
\nonumber
&& a_1= { 93 - 10\,n_f \over 36} \ \ \ \ , \ \ \ \
 {\beta_{\!_0}} = {11 - {2\over 3 } n_f }  \ \ ,
\end{eqnarray}
where
\begin{eqnarray}
\nonumber
&& c_{\,0}(t) \ \ {\mathop =_{ka \ll 1}} \ \ {{\beta_{\!_0}} \,
\alpha_{\hskip-1.pt s} (t)\over 4\,\pi}
\,\left[ \ln {t^{1/2}\,a\over 2} - 1 - 2\,{\gamma_{\!_E}}
+ {(ka)^2 \over 12} + {(ka)^4 \over 40} + \dots \right] \ ,\\
\nonumber
&& k\equiv m v\ \ , \ \
a \equiv { 2 / m {C_{\!_F}} \widetilde{\alpha}_{\hskip-1.3pt s}(t) }\ .
\end{eqnarray}
An exact expression for $c_0(t)$ (which represents the
radiative corrections) may be found in Ref. \cite{bb:kassa}.
Eq. (4.6) includes also nonperturbative corrections with
$\ \langle \alpha_{\hskip-1.pt s} G^2\rangle  = 0.042 \pm 0.020\ GeV^4$,
and the evaluation is valid until these are of order unity, $i.e.$,
down to a critical velocity
\begin{equation}
v_{crit} \sim \left( { 2\, \pi \langle \alpha_{\hskip-1.pt s} \,G^2\rangle
\over 192 \,{\beta_{\!_0}} \, m^4}
\right)^{1/6} \ \ \ .
\eqnum{4.7}
\end{equation}
For the heavy quarks, we find it more convenient
to reexpress the $\overline{m}$
in terms of the {\em pole} masses $m$: one has$^{\cite{bb:coquereaux}}$
\begin{eqnarray}
\eqnum{4.8}
&& {{\overline{m}}}(t) = m\,\left[ { \alpha_{\hskip-1.pt s}(t)
\over \alpha_{\hskip-1.pt s}(m^2) } \right]^{-\gamma_0/\beta_{\!_0}}\;
\left\{
  1 + { A\,\alpha_{\hskip-1.pt s}(t)
- ({C_{\!_F}}-A)\alpha_{\hskip-1.pt s}(m^2) \over \pi}
\right\} \ \ ,\\
\nonumber
&& A \equiv
{ {\beta_{\!_1}} \gamma_0 - {\beta_{\!_0}} \gamma_1 \over {\beta_{\!_0}}^2}\ ,
\ \ \ \gamma_0 = -4 \ ,\ \ \ \ \beta_{\!_1} = 102 - {38\over 3}\,n_f\ ,
\ \ \ \ \gamma_1 = -{202\over 3} + {20\over 9} \, n_f \ \ ,
\end{eqnarray}
with the values of the pole masses given by$^{\cite{bb:titard}}$
\begin{equation}
m_c = 1570 \pm 60 \ MeV \ \ \ , \ \ \
m_b = 4906 \pm 85 \ MeV\ \ ,
\eqnum{4.9}
\end{equation}
and we take$^{\cite{bb:mtop}}$
\begin{equation}
m_t = 180 \pm 12 \ GeV \ \ .
\eqnum{4.10}
\end{equation}
A crude estimate can now be obtained if one uses for $R_q(t) \ (q=c,b,t)$
the following approximation:
\begin{equation}
R_q^{(I)}(t) = \left\{
   \begin{array}{lcl}
      R_q^{\rm{l.e.}} \ & \ {\rm {if}}\  & \ v_{crit} < v \le 1/2 \\
      R_q^{\rm{h.e.}} \ & \ {\rm {if}}\  & \ v > 1/2
   \end{array}
\right.
\eqnum{4.11}
\end{equation}
The expression is however discontinuous at $v=1/2$. To obtain a smooth
joining of the low and high energy regions, we can use the following
two interpolation curves for $R_q(t)$:
\begin{eqnarray}
\eqnum{4.12}
&& R_q^{(II)}(t) = (1-v)  R_q^{\rm{l.e.}} + v\; R_q^{\rm{h.e.}} \ ,\\
\eqnum{4.13}
&& R_q^{(III)}(t) = (1-v^3)  R_q^{\rm{l.e.}} + v^3\; R_q^{\rm{h.e.}} \ .
\end{eqnarray}
We consider $R_q^{(II)}$ to be a very reasonable approximation, and
take the difference between $R_q^{(I,II,III)}$ as
an estimate of the systematic error due interpolations.
We also take into account the error due to the
input parameters
$\alpha_s(m_{\tau})$, $\alpha_s(M_{Z})$, and $m_t$, as well as the
error due to the non-perturbative contribution.
The last is obtained by setting
$\langle \alpha_s G^2 \rangle =0$ in Eq. (4.6)
and allowing the velocity $v$ of
the heavy quarks to go to zero.
Our best estimate for the contribution of the regions
in the continuum is:
\begin{eqnarray}
\eqnum{4.14}
&& a_h ( {\rm{continuum}}, t^{1/2} > \sqrt{2} \; GeV) =
(826.4 \pm 7.8 \pm 9.8 \pm 8) \times 10^{-11} \ ,\\
\eqnum{4.15}
&& \Delta \alpha_h ({\rm{continuum}}, t^{1/2} > \sqrt{2} \; GeV) =
	(226.6 \pm 3.7 \pm 1 \pm 1) \times 10^{-4}\ .
\end{eqnarray}
The first systematic error is due to interpolations, the second is due
to the error in the input parameters, and the last is the
non-perturbative one.

% %%%%%%%%%%%%%%%%%%%%%%%%%%%%%%%%%%%%%%%%%%%%%%%%%%%%%%%%%%%%%%%%%%%%%%%%
\section{Conclusion}
\label{sec:5}
% %%%%%%%%%%%%%%%%%%%%%%%%%%%%%%%%%%%%%%%%%%%%%%%%%%%%%%%%%%%%%%%%%%%%%%%%

Besides using our theoretical estimates for the high energy regions
\begin{equation}
t^{1/2}>1.1\,{\rm GeV},\;\sqrt{2}\,{\rm GeV}
\end{equation}
for ${\rm Im}\Pi(t)$ an extra (slight) improvement may be incorporated
if we repeat the dispersive analysis of Ref.{\cite{bb:casas}}
in the $\rho$ region using recent data. The reason
for this improvement is the following: in that paper the $\rho$
contribution to (say) $a_h(t^{1/2}<0.9\,{\rm GeV})$, had been estimated
from fits to the pion form factor $F_{\pi}$ alone (but both in the
spacelike and timelike regions) or imposing also the values of the
$\pi\pi$ phase shifts.
 The first method gave $m_{\rho}=768\,{\rm MeV}$, the second
 $m_{\rho}=778\pm 2\,{\rm MeV}$, both compatible (at the $2\sigma$
level) with the then preferred experimental value $m_{\rho}=769\pm
3\,{\rm MeV}$.  Although, as explained in Ref.{\cite{bb:casas}},
the first method was considered more reliable, both were combined
taking the difference between the two determinations as a measure of
the {\it systematic} error of the calculation. Since the presently
accepted experimental figure, $m_{\rho}=768.1\pm 0.5\,{\rm MeV}$
clearly discriminates in favor of the method based on $F_{\pi}$ only,
we can dispense with the (poorly known) $\pi\pi$ phase shifts and
avoid the systematic error.
We need only alter the evaluation by taking into account the change in
the accepted value for the $\rho$ width,
$\Gamma_\rho=151.5\pm1.2\,{\rm MeV}$ from that used in Ref.{\cite{bb:casas}},
$158\,{\rm MeV}$ (compatible with the 1985 experimental value,
$\Gamma_\rho=154\pm 5\,{\rm MeV}$ but not with the presently preferred
one). To first order this is easily taken
into account as a variation $\Delta\Gamma/\pi m_\rho=-0.27\,\%$.

The results are summarized in the following tables where we also
report, for purposes of comparison, the evaluations of the first (EJ)
and third (MZ) papers of Ref.\cite{bb:eidelman}.
We have composed quadratically statistical and
systematic errors: there are so many of the last, and of such varied
origins, that they may be taken to behave statistically on the
average. For the muon anomaly we have,
\begin{eqnarray}
\nonumber
a_h(t^{1/2}<\sqrt{2})=6200.0\pm 101.4\times 10^{-11} &&
\ \ \ ({\rm This \;work})\\
\nonumber
a_h(t^{1/2}<\sqrt{2})=6342.3\pm 137.7\times 10^{-11} &&
\ \ \ ({\rm EJ,\;Ref.\cite{bb:eidelman}})
\end{eqnarray}
and
\begin{eqnarray}
\nonumber
a_h(t^{1/2}>\sqrt{2})=913.4\pm 14.9\times 10^{-11} &&
\ \ \  ({\rm This \;work})\\
\nonumber
a_h(t^{1/2}>\sqrt{2})=908.2\pm 77.1\times 10^{-11} &&
\ \ \  ({\rm EJ,\;Ref.\cite{bb:eidelman}})\,,
\end{eqnarray}
with the overall results
\begin{eqnarray}
\nonumber
a_h=7113.4\pm 102.5\times 10^{-11} &&
\ \ \  ({\rm This \;work})\\
\nonumber
a_h=7250.4\pm 157.6\times 10^{-11} &&
\ \ \  ({\rm EJ,\;Ref.\cite{bb:eidelman}})\,.
\end{eqnarray}
which agree within the quoted errors.

For the running QED coupling, one has
\begin{eqnarray}
\nonumber
\Delta\alpha_h(t^{1/2}<\sqrt{2})=50.3\pm 1.6\times10^{-4} &&
\ \ \  ({\rm This\;work})\\
\nonumber
\Delta\alpha_h(t^{1/2}<\sqrt{2})=47.92\pm 1.06\times10^{-4} &&
\ \ \  ({\rm EJ,\;Ref.\cite{bb:eidelman}})\,,
\end{eqnarray}
and
\begin{eqnarray}
\nonumber
\Delta\alpha_h(t^{1/2}>\sqrt{2})=239.12\pm 4.1\times10^{-4} &&
\ \ \  ({\rm This\;work})\\
\nonumber
\Delta\alpha_h(t^{1/2}>\sqrt{2})=232.45\pm 6.45\times10^{-4} &&
\ \ \  ({\rm EJ,\;Ref.\cite{bb:eidelman}})\,,
\end{eqnarray}
and now the full results are
\begin{eqnarray}
\nonumber
\Delta\alpha_h=289.42\pm 4.35\times10^{-4} &&
\ \ \  ({\rm This\;work})\\
\nonumber
\Delta\alpha_h=280.37\pm 6.54\times10^{-4} &&
\ \ \  ({\rm EJ,\;Ref.\cite{bb:eidelman}})\,.
\end{eqnarray}
Our central value for $\Delta\alpha_h$ is slightly higher than
that of Ref.\cite{bb:eidelman} (EJ), but deviates at the 3-4 $\sigma$--level
from that given by the third article of Ref.\cite{bb:eidelman}:
\begin{eqnarray}
\nonumber
\Delta\alpha_h=273.2\pm 4.2\times10^{-4} &&
\ \ \  ({\rm MZ,\;Ref.\cite{bb:eidelman}})\ .
\end{eqnarray}

We conclude this paper with a few comments on possible ways to improve
the results. Certainly, better knowledge of $\alpha_s$ both on the
$\tau$ and $Z$ masses would increase the precision of the high energy
evaluations; but most of the error comes from the region $0.8\leq
t^{1/2}\leq \sqrt{2}$ GeV. Even a modest improvement of a factor two
in the error in the region between 0.8 and 1.1 GeV would result in a
substantial decrease, roughly by the same amount, of the overall error
for $a_h$, and about 20\% for $\Delta\alpha_{\rm QED}$. This emphasizes
the interest of some of the
 accelerators, projected or in construction, with the capability to
explore these energy ranges.

% %%%%%%%%%%%%%%%%%%%%%%%%%%%%%%%%%%%%%%%%%%%%%%%%%%%%%%%%%%%%%%%%%%%%%%%%
% 			{\bf References}
% %%%%%%%%%%%%%%%%%%%%%%%%%%%%%%%%%%%%%%%%%%%%%%%%%%%%%%%%%%%%%%%%%%%%%%%%


\begin{thebibliography}{99}
% ref 1
\bibitem{bb:eidelman}
S.~Eidelman and F.~Jegerlehner, PSI--PR--95--1, 1995;\\
M.~L.~Swartz, SLAC--PUB--6710, 1995;\\
A.~D.~ Martin and D.~Zeppenfeld, MAD/PH/855, 1994.
%
% ref 2
% \bibitem{bb:brookheaven}
% Brookheaven.
%
% ref 3
% \bibitem{bb:electroweak}
% Electroweak.
%
% ref 4
\bibitem{bb:casas}
J.~A.~Casas, C.~L\'opez, and F.~J.~Yndur\'ain,
{Phys. Rev.} {\bf D$\;$32}, 736 (1985).
%
% ref 5
\bibitem{bb:kassa}
K.~Adel and F.~J.~Yndur\'ain, FTUAM 95--2, hep-ph/9502290, 1995.
%
% ref 6
\bibitem{bb:chetyrkin}
K.~G.~Chetyrkin, J.~H.~K\"uhn,
{ Phys. Lett.}, {\bf B$\;$248}, 359 (1990);\\
%
S.~G.~Gorishny, A.~L.~ Kataev, S.~A.~Larin and L.~R.~Sugurladze,
{ Phys. Rev.} {\bf D$\;$43},\\
%
1633 (1991). K.~G.~Chetyrkin, A.~L.~Kataev and F.~V.~Tkachov,
{ Phys. Lett.} {\bf B$\;$85}, 277 (1979); \\
%
M.~Dine and J.~Sapiristein, { Phys. Rev. Lett.} {\bf 43}, 668 (1979);\\
%
W.~Celmaster and R.~J.~Gonsalves, { Phys. Rev. Lett.} {\bf 44}, 560 (1980);
\\
%
S.~G.~Gorishny, A.~L.~ Kataev and S.~A.~Larin,
{ Phys. Lett.} {\bf B$\;$259}, 144 (1991).
%
% ref 7
\bibitem{bb:pich}
See for example A.~Pich, Invited Talk at the 1994 QCD Workshop,\\
Monpellier, hep-ph/9412273.
%
% ref 8
\bibitem{bb:pdb}
Particle Data Group, { Phys. Rev.} {\bf D$\;$50}, 1175 (1994).
%
% ref 9
\bibitem{bb:yndurain}
See e.g. F.~J.~Yndur\'ain,
{\em The Theory of Quark and Gluon Interactions}, Springer, (1993),\\
and work quoted there.
%
% ref 10
\bibitem{bb:coquereaux}
R.~Coquereaux, {Phys. Rev.} {\bf D$\;$23}, 1365 (1981); \\
D.~J.~Broadhurst et al., Z. Phys. {\bf C$\;$48}, 673 (1990).
% D.~J.~Broadhurst et al., Preprint MPI--PhT/94--2 (1994).
%
% ref 11
\bibitem{bb:titard}
S.~Titard and F.~J.~Yndur\'ain, { Phys. Rev.} {\bf D$\;$49}, 6067 (1994);\\
{ Phys. Rev.} {\bf D$\;$51}, 6348 (1995).
%
% ref 12
\bibitem{bb:mtop}
CDF collaboration, Phys. Rev. Lett. 74, 2626 (1995);\\
D0 collaboration, Phys. Rev. Lett. 74, 2632 (1995).
%
\end{thebibliography}
\end{document}